# Driving frequency effect on discharge parameters and higher harmonic generation in capacitive discharges at constant power densities

Sarveshwar Sharma[1], Nishant Sirse[2], Animesh Kuley[3], Abhijit Sen[1] and Miles M Turner[4]

[1]Institute for Plasma Research (IPR) and HBNI, Gandhinagar 382428, India

[2]Institute of Science & Laboratory Education, IPS Academy, Indore 452012, India

[3]Department of Physics, Indian Institute of Science, Bangalore 560012, India

[4]School of Physical Sciences and NCPST, Dublin City University, Dublin 9, Ireland

E-mail: nishantsirse@ipsacademy.org

**Abstract**

Very high frequency (VHF) driven capacitive discharges are now being increasingly adopted for plasma-based materials processing due to their high processing rates and lower substrate damage. Past studies related to complex plasma dynamics and higher harmonics generation in such systems were limited to constant voltage/current conditions, whereas, industrial systems are mostly driven by constant power density sources. In the present study, using particle-in-cell (PIC) simulation, we explore the dynamics of collisionless symmetric capacitive discharges that is operated at constant power densities. Our focus is on the effect of the driving frequency on the discharge parameters like the *electron density/temperature, the electron energy distribution function* (EEDF), the *ion energy distribution function* (IEDF), and the generation of higher harmonics in the device. The simulations are performed for a driving frequency from 27.12-100 MHz in argon plasma at a gas pressure of 1 Pa and for two values of the power density, namely, 2 kW/$m^3$ and 20 kW/$m^3$. It is observed that the required discharge voltage for maintaining constant power density decreases and discharge current increases with an increase in the driving frequency. A transition frequency is observed at both power densities. The density decreases (electron temperature increases) before the transition frequency and the trend is reversed after crossing the transition frequency. The EEDF shows an enhancement in the population of the mid-energy range of electrons as the driving frequency increases up to the transition frequency thereby changing the shape of EEDF from bi-Maxwellian to nearly Maxwellian, and then transforms into a nearly bi-Maxwellian at higher driving frequencies. The IEDF at the electrode surface shows bimodal behaviour at a lower driving frequency, becoming more pronounced at a power density of 20 kW/$m^3$, and then turning into a single energy peak. The corresponding maximum ion energy is found to decrease with driving frequency.

## 1. Introduction

Plasma based techniques for materials processing, such as in thin film depositions and etching, are becoming indispensable in microelectronic device fabrications [1]. Low pressure radio frequency (RF) driven capacitively coupled plasma (CCP) discharges are mostly utilized for this purpose. In such systems, a high plasma density and a low ion bombardment energy are preferred for achieving high processing rates and minimal substrate damages, respectively. It has been observed that an increase in the driving frequency of CCP systems bestows several advantages over conventional 13.56 MHz discharges. In particular, an enhanced plasma density and a lower self-bias can be generated for a very high frequency (VHF) plasma excitation. A dual frequency CCP consisting of a simultaneous application of a VHF and a low frequency driver has often been proposed for an independent control of ion flux and ion energy [2]. In such systems, the VHF power is used to control the plasma density in the discharge thereby controlling the ion flux towards the substrate [3]. VHF plasma operation is further suited for the generation of unique gas-phase chemistry. This is due to a transition in the Electron Energy Distribution Function (EEDF) as a function of the driving frequency [4-5] that leads to different excitation, dissociation, and ionization processes. Despite several advantages, the power deposition mechanisms in such systems are not very well understood due to their complex non-linear behaviour. Thus, the investigation of plasma heating and discharge parameters in VHF excited CCPs are extremely important for further optimization of process conditions and better control strategies for the next generation advanced processing reactors.

Recent studies performed in VHF CPP systems have shown power depositions through the generation of an energetic electron beam from the proximity of the sheath edge [5-7]. At a low gas pressure, such an electron beam can travel without collisions through the bulk plasma and interact with the opposite moving sheath edge. Depending on the driving frequency, the energetic electrons can be confined within the discharge system and thereby act as a possible mechanism for an enhancement in the plasma density in the VHF regime. The electron beam can also be responsible for the generation of electric field non-linearities and higher harmonics in the bulk plasma and thereby cause heating of the low energy electrons [8-10]. This heating effect can lead to a significant change in the EEDF. Multiple electron beams have also been reported in some of the past studies of VHF CCPs [5,7]. Berger et al [11] experimentally showed the formation of multiple electron beams within a single phase of the sheath expansion using phase resolved optical emission spectroscopy. A combination of discharge voltage and driving frequency was also proposed to achieve an independent control of ion flux and ion energy in a CCP system [12]. It should be noted that most of the above studies were performed for voltage driven CCPs. However, industrial plasma processing reactor environments are mostly defined and controlled by the RF power level. Therefore, it is important to reproduce such studies for constant power conditions. Additionally, the study of EEDF and IEDF are highly crucial for understanding and controlling the plasma chemistry and ion energy at the surface, respectively.

Motivated by the above considerations, we have performed particle-in-cell (PIC) simulations to investigate the effect of driving frequency on the electron heating mechanism and various plasma parameters in a CCP system at constant power densities. In the past, there have been very few theoretical and/or experimental studies devoted to CCP behaviour under constant power density conditions. Yan and Goedheer [13] studied the behaviour of VHF CCP discharges in a mixture of silane and hydrogen at pressures below 300 mTorr and for frequencies from 13.56 to 65 MHz using a 2d PIC/MC code. They showed that with an increase in the driving frequency, the ion energy at the substrate decreases drastically without any modification in the ion flux. The ion energy spectrum becomes more peaked and no changes in the dissociation and ionization are observed, while vibrational excitation increases strongly. Amanatides and Mataras [14] investigated the driving frequency variation under constant power conditions in hydrogen radio frequency discharges using electrical and optical diagnostics, and employed a theoretical discharge model. Their results predicted a decrease in the discharge voltage, increase of the discharge current and a decrease of the discharge impedance. As driving frequency increases, the power consumed in ion decreases and power consumed in electron acceleration increases leading to an increase in the electron drift velocity and electron density. Meanwhile, the dissociative excitation and ionization rates are found to decrease with frequency. The experiments by Ahn et al [15] in argon capacitive discharges in the frequency range from 9 to 27.12 MHz found that the effective electron temperature decreased and the electron density either remained constant or decreased. The EEDF remained bi-Maxwellian for the entire frequency range. Zhu et al [16] measured the electron density and ion energy for the driving frequency range of 13.56 MHz to 156 MHz e in an argon discharge. Their results show an initial increase in electron density with driving frequency and saturation at higher driving frequencies. Meanwhile, the ion energy decreases with frequency but does. not agree with the trend predicted by a simple model based on a Maxwellian EEDF with no electromagnetic effects at higher driving frequencies. Abdel-Fattah [17] showed experimentally that the EEDF changes its shape from a Maxwellian at 13.56 MHz to a bi-Maxwellian in the frequency range 27-56 MHz and eventually comes to a Maxwellian at frequencies above 76 MHz. The electron density and temperature peak in the same frequency range and decrease as the driving frequency increases. The above studies have shown not shown any definitive universal trend and therefore a comprehensive study is required to understand the plasma behaviour as a function of the driving frequency under constant power density conditions. In addition, one also needs to gather information about the generation of higher harmonics and the nature of the electric field non-linearity under a constant power driving.

Our paper is organized as follows. In section 2, we provide a description of the simulation scheme, initial condition set-up and discharge parameters considered in this work. The physical interpretation and discussion of the simulation results are presented in section 3. Finally, a brief summary and some concluding remarks are given in section 4.

## 2. Simulation scheme and parameters

The current analysis is based on a self-consistent 1D3V particle-in-cell (PIC)/Monte-Carlo Collisions (MCC) simulation scheme using a well-tested and benchmarked code [18] that has been previously utilized to study the effects arising from varying the driving frequency in CCP discharges [5, 8-10, 12, 19]. In the present study, the electrodes are assumed to be planar, infinite dimensional and placed parallel to each other with a separation of 3.2cm. The powered electrode is driven by a voltage with a sinusoidal waveform, $V(t) = V_0 sin(2\pi f t)$, where $V_0$ is the amplitude, and $f$ is the frequency of the applied RF. The other electrode is grounded. An appropriate choice of the spatial step size (i.e. smaller than the Debye length) and temporal step size (to resolve the electron plasma frequency) have been made to take care of the accuracy and stability of the numerical computations. All the important particle-particle interactions like the electron-neutral (elastic, inelastic and ionization) and the ion-neutral (elastic, inelastic and charge exchange) collisions are taken into account for all sets of simulations. However, processes like multi-step ionization, metastable pooling, partial de-excitation, super elastic collisions and further de-excitation are not considered here for simplification. The species, reactions and cross sections used here are taken from well-tested sources listed in the references [20-21]. In the simulation, the production of metastables (i.e.

$Ar$*, $Ar$**) has been considered though we did not track them for output diagnostics. Both the electrodes are perfectly absorbing for electrons and ions and the secondary electron emission is ignored for the present case. The simulation region is divided by a grid into 512 cells and is cell is populated by 100 particles. The background neutral gas is distributed uniformly with a temperature similar to that of ions i.e. 300 K. The simulations were run for more than 5000 RF cycles (~7000 RF cycles at 100 MHz) to achieve steady state solutions.

## 3. Results and discussions

Figure 1 shows the effects of the driving frequency on the discharge voltage at constant power densities of ~2 kW/m$^3$ and ~20 kW/m$^3$. The total power density is the sum of averaged ion and electron power densities over one RF period and the discharge gap i.e. $\{< j_i(x,t)E(x,t) > + < j_e(x,t)E(x,t) >\}$, where, $j_i$ and $j_e$ are the ion and electron current densities respectively, $E$ is the electric field and <..> denotes time averaging over one RF period and space averaging over the discharge gap. The driving frequency is varied from 27.12 MHz to 100 MHz. As shown in figure 1, the discharge voltage decreases with an increase in the driving frequency in order to maintain a constant power level. At 2 kW/m$^3$, the discharge voltage decreases from ~250 V at 27.12 MHz to 60 V at 100 MHz, whereas, at 20 kW/m$^3$, it decreases from ~800 V at 27.12 MHz to 230 V at 100 MHz. The best fit to the data indicated by the solid curves point towards an approximate scaling law given by $V_{rf} \propto f^{-1}$ i.e. the discharge voltage varies as the inverse of the driving frequency. Such a decrease in the discharge voltage is in agreement with results from previous studies [22-24] and can be attributed to a change in the electron and ion power dissipation when the driving frequency is varied. It is observed that the electron power absorption increases, whereas, the ion power absorption decreases with a rise in the driving frequency. This leads to a decrease in the discharge impedance, and consequently, a rise in the discharge current from ~ 20 A/m$^2$ at 27.12 MHz to ~50 A/m$^2$ at 100 MHz for 2 kW/m$^3$ and from ~ 68 A/m$^2$ at 27.12 MHz to ~218 A/m$^2$ for 20 kW/m$^3$. Therefore, the required discharge voltage has to decrease to maintain a constant power density.

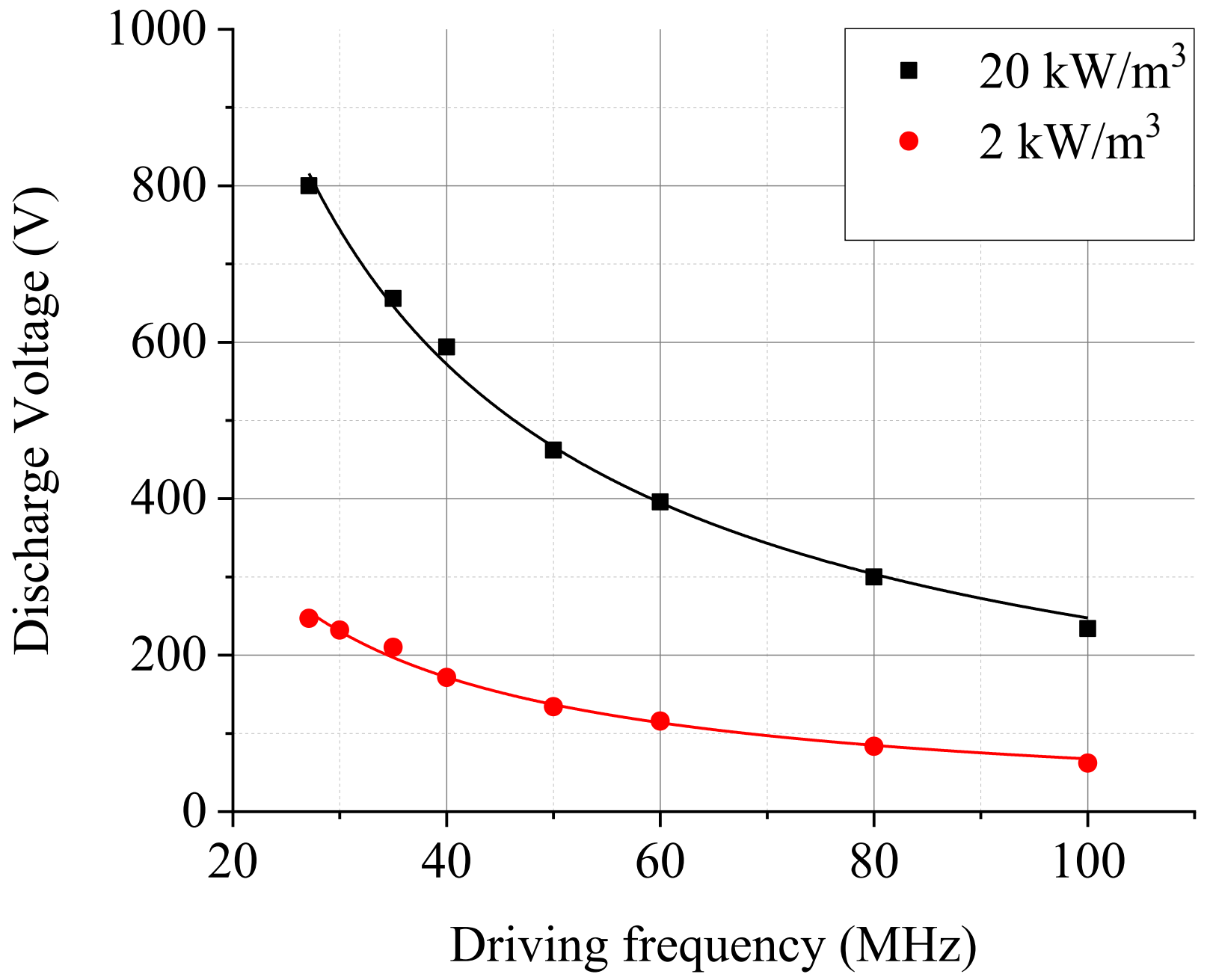

**Figure 1.** Discharge voltage versus driving frequency at power densities of 2 kW/m$^3$ and 20 kW/m$^3$. The solid curves fitted to the data have an inverse frequency ($f^{-1}$) dependence.

Figure 2 (a) and 2 (b) show the effects of the driving frequency on the electron density ($n_e$) and the electron temperature ($T_e$) for constant power densities of 2 kW/m$^3$ and 20 kW/m$^3$, respectively, where the averaged values of $n_e$ and $T_e$ over the discharge gap have been plotted. The variation of $n_e$ and $T_e$ values as a function of the frequency show some interesting features. In the low frequency regime the density initially decreases before showing a rising trend in figure 2 (a). The temperature shows a reverse trend in the same regime – it rises before showing a decreasing trend. At much higher values of the frequency, as one approaches 100 MHz, the density appears to saturate with a slight decreasing trend towards the end while the temperature shows a saturation followed by a slight rising trend in the same region. The same features are also seen in figure 2(b) but at this higher power operational regime the second regime of density saturation is not seen and perhaps occurs for a higher frequency value that has not been explored numerically. At 2 kW/m$^3$, $n_e$ is in the range of ~$1\times10^{15}$ m$^{-3}$ to ~$2.25\times10^{15}$ m$^{-3}$, and for 20 kW/m$^3$ it is in the range of ~$8\times10^{15}$ m$^{-3}$ to ~$2\times10^{16}$ m$^{-3}$ in the frequency interval of 40 MHz to 80 MHz. The value of $T_e$ varies in the range of ~ 3.25 eV to 2.5 eV at 2 kW/m$^3$, and from ~2.8 eV to 2 eV at 20 kW/m$^3$. So the general trend is an initial decrease in $n_e$ up to a transition frequency followed by an increase with a further increase in driving frequency, and then a near saturation at higher driving frequencies. On the other hand, the $T_e$ first increases up to the transition frequency and then decreases with a further rise in driving frequency, and finally approaches saturation at higher driving frequencies. The transition frequency is ~35 MHz at 2 kW/m$^3$ and ~ 40 MHz at 20 kW/m$^3$ power density. The variation of the density versus the driving frequency qualitatively agrees with the experimental results of Zhu et al [16] (conducted for the driving frequency range from 13.56 to 156 MHz) except that they did not observe an initial dip in $n_e$ for the lower driving frequency range. However, the experimental study of Ahn et al [15], which was performed for a lower, and limited range of driving frequencies, namely, from 9 to 27.12 MHz, showed that the electron density decreases and the electron temperature increases. Ahn et al [15] further showed that the EEDF remains bi-maxwellian for the entire driving frequency range, whereas, an increase in the collisional power absorption enhances the low energy electron heating responsible for the increasing trend of the electron temperature. For our simulation results, we show by analysing the EEDF that the low energy electrons are heated, up to the transition frequency, due to the presence of electric field transients in bulk plasma generated by energetic electrons ejected from the vicinity of the expanding sheath edge.

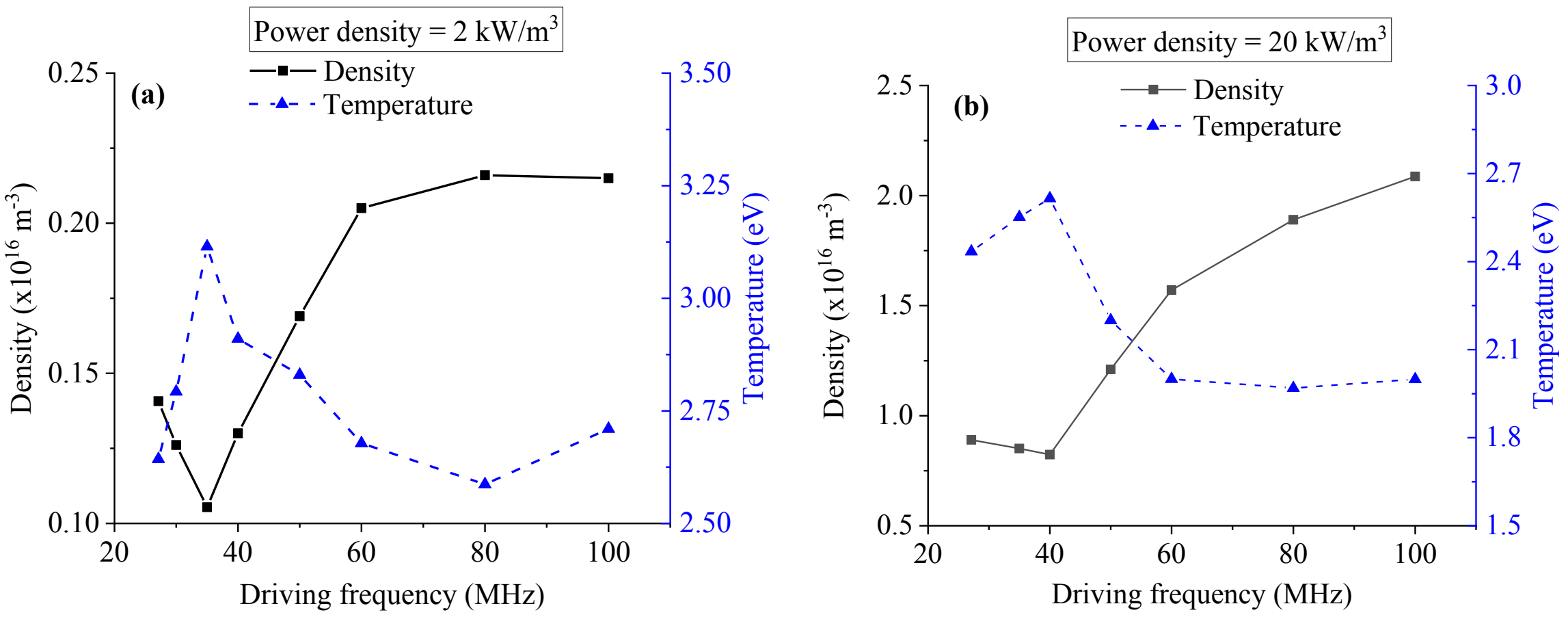

**Figure 2.** Spatial average electron density and temperature versus driving frequency at power densities of a) 2 kW/m$^3$ and b) 20 kW/m$^3$.

The increasing trend of $T_e$ up to the transition frequency and concurrent decreasing trend of $n_e$ in order to maintain the power balance are associated with the heating of the low energy electrons that are confined in the bulk plasma. This could be observed in the electron energy distribution function (EEDF). Figure 3 (a) and 3 (b) show the EEDF plotted for different frequencies and power densities. Three frequencies are displayed in each figure: 1) 27.12 MHz, 2) Transition frequency (35 MHz @ 2 kW/m$^3$ and 40 MHz @ 20 kW/m$^3$), and 3) 100 MHz. As shown in figure 3 (a), the shape of EEDF at 27.12 MHz is strongly bi-Maxwellian with a large population of low energy electrons (0.5 *eV*). As the driving frequency increases to transition frequency, low energy electrons are heated and start diffusing from the low energy region to the high energy region of EEDF. This changes the shape of EEDF to one that is nearly Maxwellian. As the driving frequency increases to 100 MHz, the population of both low energy and high energy electrons increases. However, the low energy electron population increases at a higher rate and induces the EEDF shape to become nearly Maxwellian. As shown in figure 3 (b), similar results are observed at a power density of 20 kW/m$^3$ except that at 100 MHz the EEDF shape is nearly bi-Maxwellian. The EEDF transition from 27.12 MHz up to the transition frequency explains an increase in the electron temperature. Ahn et al [15] explained the heating of low energy electrons based on the calculation of collisional (Ohmic) power absorption per unit area in the bulk plasma ($S_{\mathrm{Ohm}}$) using a homogeneous plasma model. In the present study, the collisional heating is negligible in the plasma bulk since the electron mean free path is higher than the bulk plasma length for different operating conditions considered here. Therefore, the mechanism of EEDF transition is due to a different heating mechanism present in the discharge. This is discussed in the next paragraph.

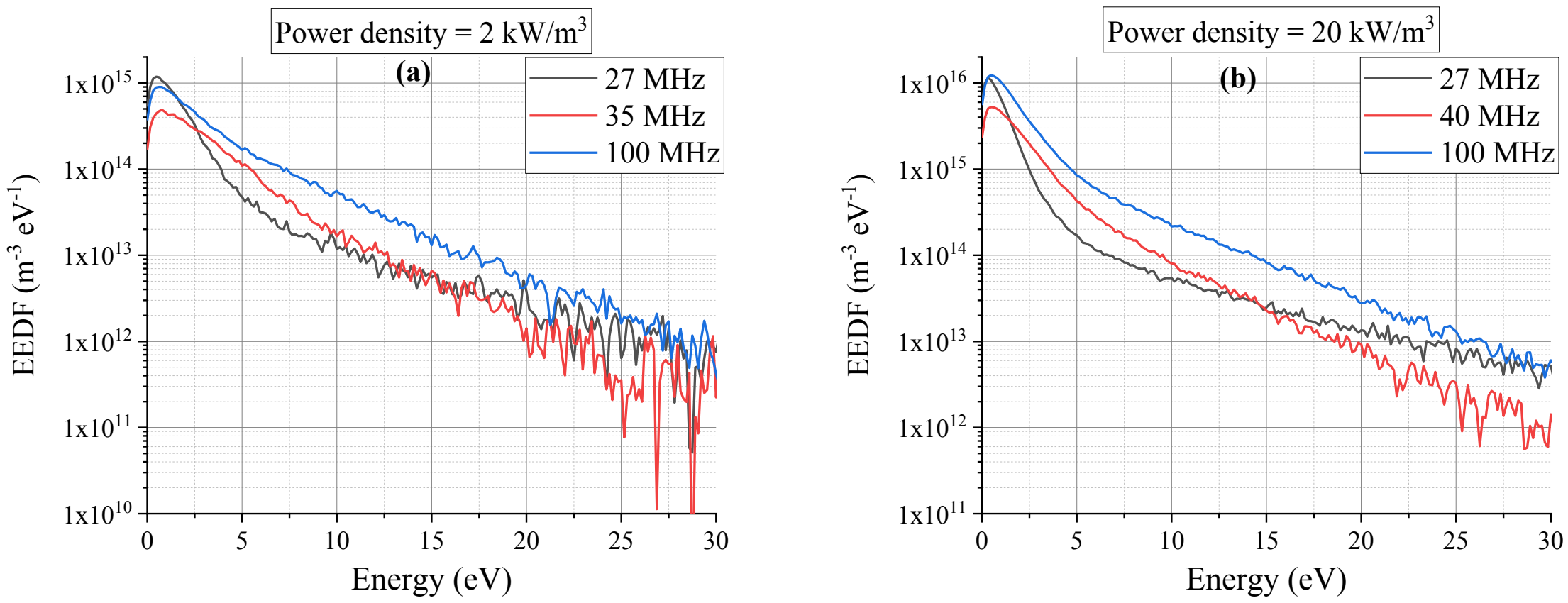

**Figure 3.** Electron Energy Distribution Function (EEDF) for different driving frequencies at power densities of a) 2 kW/m$^3$ and b) 20 kW/m$^3$.

To elucidate the heating mechanism and transition in the shape of the EEDF, we have examined the spatio-temporal evolution of the electric field and the time-averaged electron heating. Figure 4 shows the spatio-temporal evolution of the electric field averaged over the last hundreds of RF cycles and plotted over an interval of 2 RF periods. The corresponding electric field at the centre of the discharge and its Fast Fourier Transform (FFT) are presented in figure 5. As can be seen in figure 4 (a), at 27.12 MHz, the electric field is mostly confined in the sheath region and is nearly zero in the plasma bulk. The electric field at the centre of the discharge and its FFT (figure 5 (a)) show the power in the fundamental mode of oscillation, which is approximately 22%. The harmonics (> $16^{th}$) above the electron plasma frequency (~450 MHz) at the centre of the discharge are significantly lower i.e. below 5%. The ambipolar electric field keeps the low energy bulk electrons confined inside the bulk plasma and heating of the high energy electrons through interaction with the sheath edge results in a strongly bi-Maxwellian EEDF. As the driving frequency approaches the transition frequency, 35 MHz at 2 $kW/m^3$, the non-linearity/higher harmonics (figure 5 (b)) starts showing up in the bulk plasma. The power in the fundamental frequency of electric field at the centre of the discharge reaches up to ~38% of the total power (figure 5 (b)) and the dominant $11^{th}$ harmonic (385 MHz) is observed to have a power value close to 20%. The electron plasma frequency in this case is ~380 MHz. It is further noticed that the high frequency transients are not able to reach up to the opposite sheath edge and therefore their confinement is not effective. This drives the discharge into a low density mode. However, the high energy electrons responsible for these electric field transients are able to redistribute their energy with low energy bulk electrons through non-linear interactions and thus an increase in the electron temperature and the transition in EEDF to a nearly Maxwellian form is observed up to the transition frequency. As the driving frequency further increases to 100 MHz (figure 4 (c)), strong electric field transients appear inside the bulk plasma, reaching up to the opposite sheath and modifying the instantaneous sheath edge position, similar to multiple frequency CCP discharges. The corresponding FFT in figure 5 (c) shows the fundamental power level to be above 40%. The calculated electron plasma frequency in this case is ~500 MHz that also appears in the FFT ($5^{th}$ harmonic, figure 5 (c)) and higher harmonics up to 1 GHz are observed. The confinement of high energy electrons between 2 sheaths drives the discharge to a high-density mode. The energy re-distribution between high energy electrons generated from near the sheath edge and low energy bulk electrons still occur and therefore the populations of both the low-energy (generated by ionization process) and the high-energy electrons increase as shown in figure 3. To a first order approximation, the generation of electric field transients at higher driving frequency is related to the sheath velocity that produces high energy electrons. The calculation shows that at 2 kW/$m^3$, the sheath velocity increases from ~$6\times10^5$ m/s at 27.12 MHz to ~$1.3\times10^6$ m/s at 100 MHz and therefore the transients are effective at 100 MHz driving frequency. Similar electric field transients (figure 4 (d) – 4 (f)) and higher harmonics (figure 5 (e) – 5 (f)) in the electric field at the centre of the discharge are observed at a power density of 20 kW/$m^3$. It should be noted that at 20 kW/$m^3$ the electric field transients (figure 4 (e) – 4 (f)) show a filamentary structure indicative of an enhanced non-linearity in the discharge (figure 5 (e) – 5 (f)). This can be attributed to the higher discharge voltage as has been remarked in the past [9]. Furthermore, the instantaneous sheath modulation is higher at a higher power density.

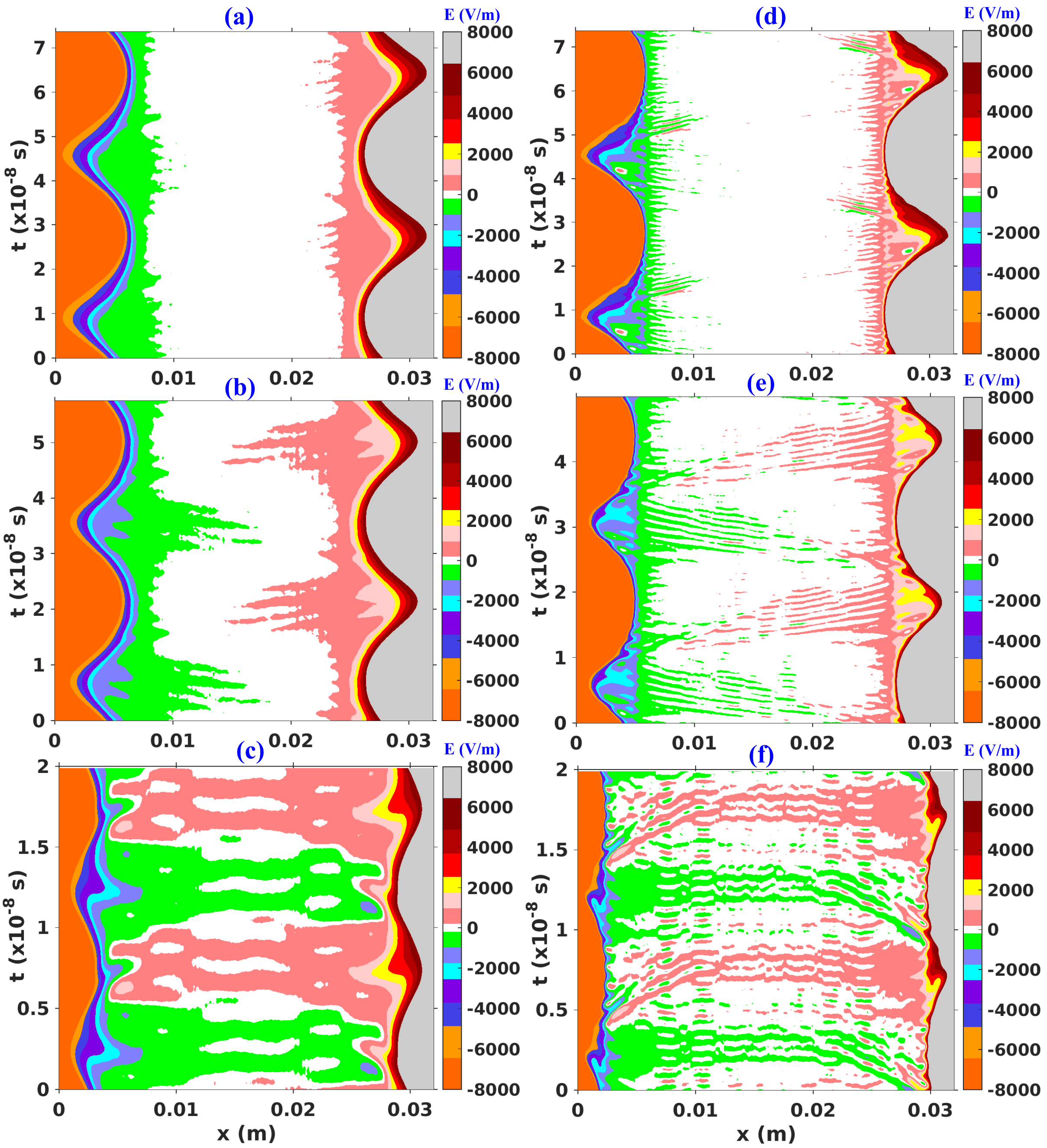

**Figure 4.** Spatio-temporal evolution of the electric field at a) 27.12 MHz, 2 kW/m$^3$, b) 35 MHz, 2 kW/m$^3$, c) 100 MHz, 2 kW/m$^3$, d) 27.12 MHz, 20 kW/m$^3$, e) 40 MHz, 20 kW/m$^3$ and f) 100 MHz, 20 kW/m$^3$.

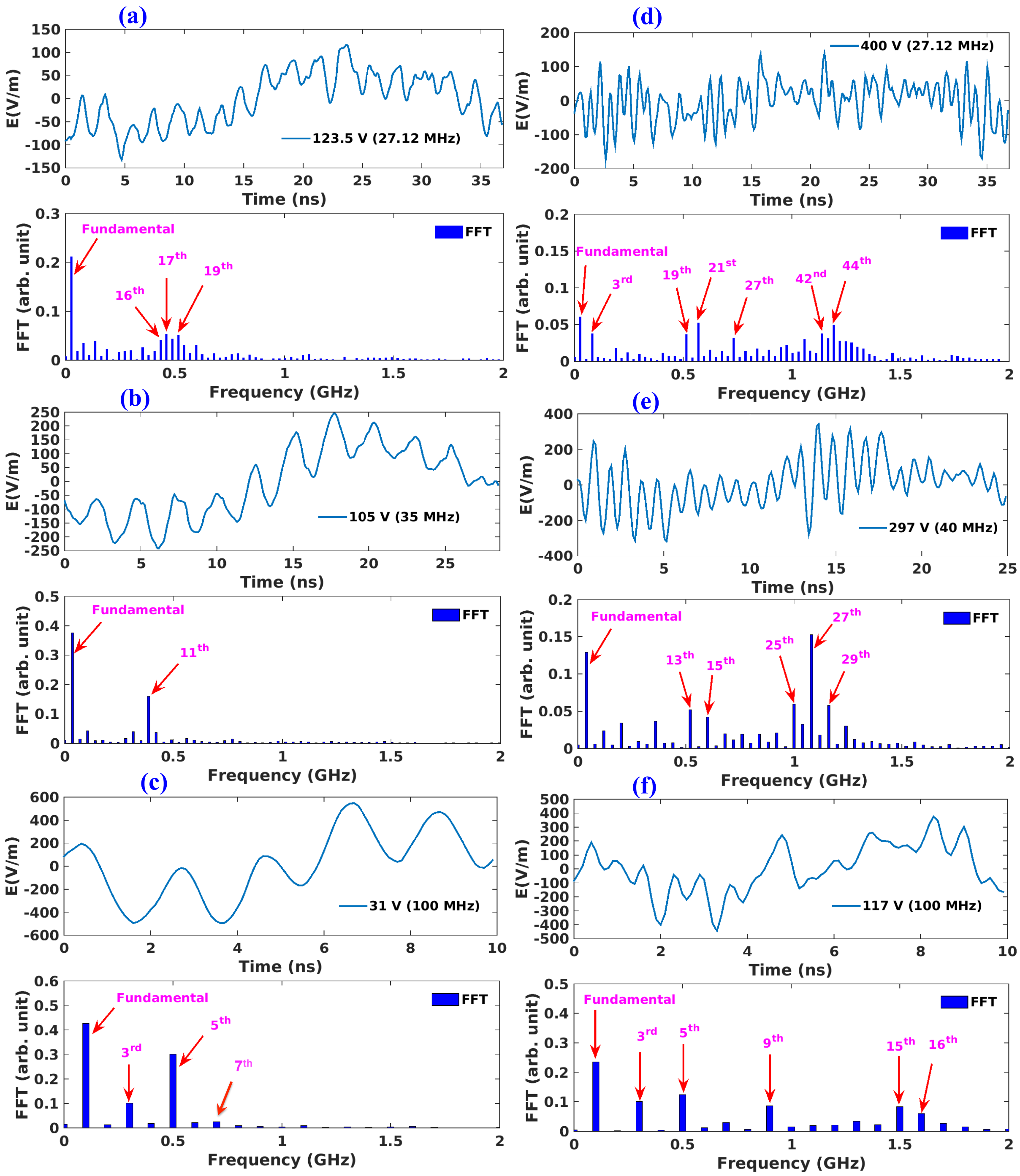

**Figure 5.** Electric field at the centre of the discharge and its Fast Fourier Transform (FFT) at a) 27.12 MHz, 2 kW/m$^3$, b) 35 MHz, 2 kW/m$^3$, c) 100 MHz, 2 kW/m$^3$, d) 27.12 MHz, 20 kW/m$^3$, e) 40 MHz, 20 kW/m$^3$ and f) 100 MHz, 20 kW/m$^3$.

Next, the effect of the driving frequency on the ion energy distribution function (IEDF) at the electrode surface is examined. Figure 6 (a) and 6 (b) show the IEDF for different driving frequencies at 2 kW/m$^3$ and 20 kW/m$^3$ respectively. As shown in figure 6 (a), at 2 kW/m$^3$, the maximum ion energy decreases from ~75 eV at 27.12 MHz to ~38 eV at 100 MHz i.e.

approximately by a factor of 2. At 20 kW/m$^3$ (figure 6 (b)), the maximum ion energy is higher in comparison to 2 kW/m$^3$ and decreases from ~190 eV at 27.12 MHz to ~70 eV at 100 MHz. At lower driving frequencies, the bimodal structure of IEDF is observed that turns into a single energy peak structure at higher driving frequencies. The bimodal structure is more pronounced at 20 kW/m$^3$. The observed behavior is attributed to the ion modulation in an RF sheath. For a collisionless RF sheath [25], the energy spread ($\Delta E_i$) is proportional to $V_s(\tau_{rf}/\tau_{ion})$ , where $V_s$ is the applied RF voltage and, $\tau_{rf}$ and $\tau_{ion}$ are the RF period and the ion transit time respectively. Based on the above expression, at a constant driving frequency ($\tau_{rf}$= constant), it can be inferred that the energy spread will increase with RF power due to a higher plasma density (lower ion transit time) and higher applied voltage. Furthermore, for a constant power density the energy spread will decrease with driving frequency due to a decrease in the RF voltage and $(\tau_{rf}/\tau_{ion})$, which is inversely proportional to the driving frequency. Hence the ion energy will transform from a bimodal structure to a single energy peak. At a constant power density, a decrease in the ion energy versus driving frequency is attributed to a decrease in the applied RF voltage and sheath width. It is observed that, for a 2 kW/m$^3$ power density, the sheath width decreases from 7 mm at 27.12 MHz to 4.5 mm at 100 MHz, and for a 20 kW/m$^3$ power density, it decreases from 6 mm (at 27.12 MHz) to 2.45 mm (at 100 MHz). The sheath width is estimated from the point where the electron sheath edge is at a maximum distance from the electrode and quasi-neutrality breaks down. A sharp dip in the electron density is observed at this point when moving from the bulk plasma towards the electrode.

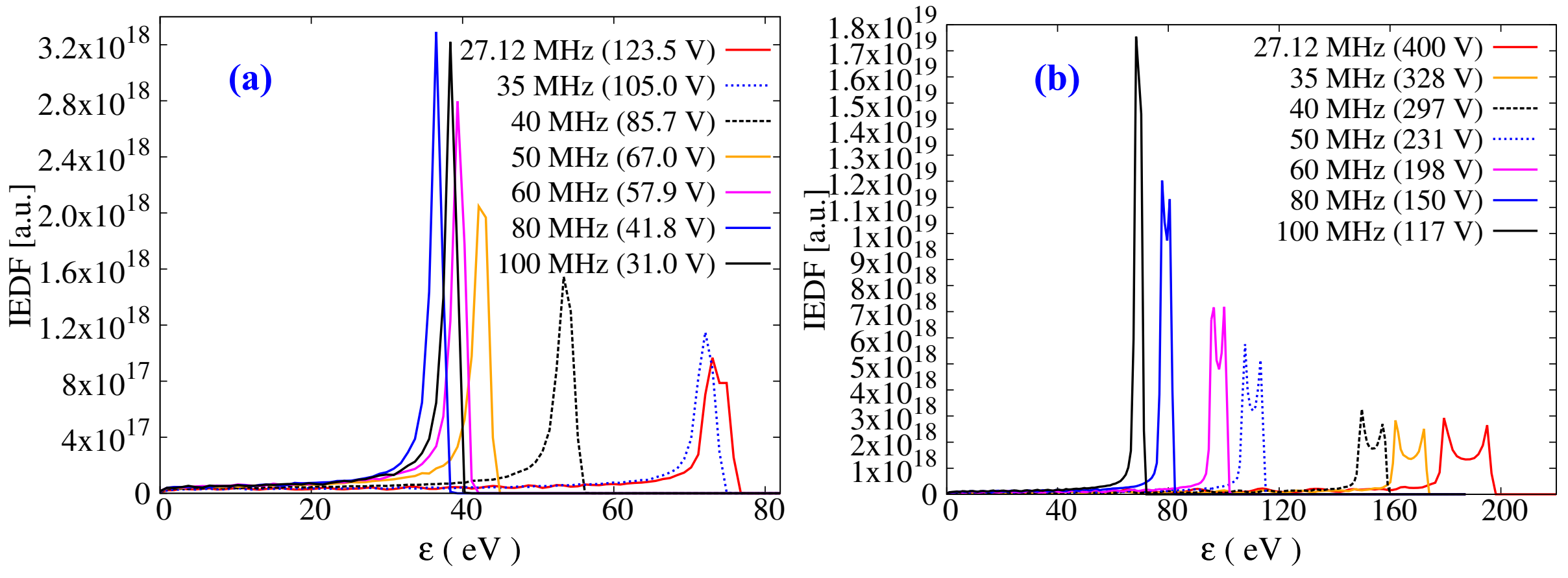

**Figure 6.** Ion Energy Distribution Function (IEDF) at the electrode surface for different driving frequencies for power densities of a) 2 kW/m$^3$ and b) 20 kW/m$^3$.

## 4. Summary and Conclusion

The effect of driving frequency on the discharge parameters of a low pressure symmetric capacitively coupled plasma discharge driven at a constant power density is studied using PIC/MCC simulations. In particular, the changes in the discharge voltage, the electron and ion energy distribution functions and the onset of higher harmonic electric field transients are examined. It is found that the discharge voltage decreases with an increase in driving frequency, which is consistent with previous studies done in the high frequency regime. However in the low frequency regime the plasma density first decreases up to a "transition frequency" and then increases with a further rise in the driving frequency reaching to a saturation at higher driving frequencies. The electron temperature shows an opposite trend. The EEDF is strongly bi-Maxwellian at lower driving frequencies, changing its shape to nearly Maxwellian at the transition frequency due to a heating of the low energy bulk electrons, and finally turning into a nearly bi-Maxwellian form at higher driving frequencies. The IEDF is bimodal at lower driving frequencies and turns into a single energy peak at higher driving frequencies. The corresponding ion energy continues to decrease with driving frequency.

An investigation of the spatio-temporal variation of the electric field within the discharge system shows that the above effect is due to the generation of electric field transients at higher driving frequencies. The Fourier spectrum of the electric field at the centre of the discharge shows no higher harmonics generation at low driving frequencies, whereas, at higher driving frequencies higher harmonics up to 2 GHz are observed. These higher harmonics are responsible for low energy bulk electron heating through non-linear interaction thereby increasing the electron temperature up to the transition frequency. As the driving frequency increases beyond the transition frequency, the transients become more energetic reaching up to the opposite sheath and modifying the instantaneous sheath edge position. Due to the confinement of high energy electrons, the ionization probability increases and therefore the plasma density increases after the transition frequency.

Our simulations highlight some important and hitherto unexplored features of plasma dynamics in a VHF CCP device that is driven at a constant power density. Among them is the existence of a transition frequency that determines when the plasma density starts increasing as a function of the driving frequency. From an operational point of view it is important to choose the driving frequency to be above this frequency to take advantage of the high density benefits of the device. The transition frequency is higher for a high-power density. The increasing trend of the density with frequency also saturates beyond a certain high frequency; hence there is no advantage in driving the device beyond this frequency regime. The ion energy continues to decrease with a single peak at a higher driving frequency and thus insures lower substrate damage. Meanwhile, the EEDF shows nearly bi-Maxwellian behaviour at higher driving frequencies and therefore the high energy processes such as dissociation, excitations etc will continue to occur with a low energy bulk electron. It must be stated that our results may not be valid for high pressure/collisional regimes or for different discharge gap values. However for the parametric regimes considered in this paper the results can provide useful operational guidelines for constant power density CCP devices. They also underscore the need for future similar explorations in other regimes in order to develop a comprehensive picture of the dynamics of VHF CCP devices over a larger operational domain.

**Acknowledgement:** Dr A Kuley is supported by the Board of Research in Nuclear Sciences (BRNS Sanctioned No. 39/14/05/2018-BRNS), Science and Engineering Research Board EMEQ program (SERB Sanctioned No. EEQ/2017/000164) and Infosys Foundation Young Investigator grant. A.S. thanks the Indian National Science Academy (INSA) for their support under the INSA Senior Scientist Fellowship scheme.

**References:**

[1] Lieberman M and Lichtenberg A J, *Principles of Plasma Discharges and Materials Processing* (Wiley: NJ, 2005)

[2] Goto H H, Lowe H D and Ohmi T 1992 *J. Vac. Sci. Technol. A* **10** 3048

[3] Boyle P C, Ellingboe A R and Turner M M 2004 *Plasma Sources Sci. Technol.* **13** 493

[4] Abdel-Fattah E and Sugai H 2003 *Jpn. J. Appl. Phys.* **42** 6569

[5] Sharma S, Sirse N, Kaw P K, Turner M M and Ellingboe A R 2016 *Phys. Plasmas* **23** 110701

[6] Rauf S, Bera K and Collins K 2010 *Plasma Sources Sci. Technol.* 19 015014

[7] Wilczek et al 2015 *Plasma Sources Sci. Technol.* 24 024002

[8] Sharma S, Sirse N, Sen A, Turner M M and Ellingboe A R 2019 *Physics of Plasmas* **26** 103508

[9] Sharma S, Sirse N, Sen A, Wu J S and Turner M M 2019 *J. Phys. D: Appl. Phys.* **52** 365201

[10] Sharma S, Sirse N, Kuley A and Turner M M 2020 *Plasma Sources Sci. and Technol.* **29** 045003

[11] Berger B, You K, Lee H-C, Mussenbrock T, Awakowicz P and Schulze J 2018 *Plasma Sources Sci. Technol.* **27** 12LT02

[12] Sharma S, Sen A, Sirse N, Turner M M and Ellingboe A R 2018 *Physics of Plasmas* **25,** 080705

[13] Yan M and Goedheer 1999 *Plasma Sources Sci. Technol.* **8** 349

[14] Amanatides E and Mataras D 2001 *Journal of Applied Physics* **89** 1556

[15] Ahn S K, You S J and Chang H Y 2006 *Appl. Phys. Lett.* **89** 161506

[16] Zhu X M, Chen W C, Zhang S, Guo Z G, Hu D W and Pu Y K 2007 *J. Phys. D: Appl. Phys.* **40** 7019

[17] Abdel-Fattah E 2013 *Vacuum* **97** 65

[18] Turner M M, Derzsi A, Donko Z, Eremin D, Kelly S J, Lafleur T and Mussenbrock T 2013 *Physics of Plasmas* **20** 013507

[19] Sharma S, Sirse N, Turner M M and Ellingboe A R 2018 *Physics of Plasmas* **25,** 063501

[20] Turner M M 2013 *Plasma Sources Sci. Technol.* 22 055001

[21] Lymberopoulos D P and Economou D J 1993 *Appl. Phys. Lett.* **63** 2478

[22] Howling et al 1992 *J. Vac. Sci. Technol. A* **10** 1080

[23] Heintze et al 1993 *J. Phys. D: Appl. Phys.* **26** 1781

[24] Kitajima et al 1998 *J. Appl. Phys.* **84** 5928

[25] Kawamura E, Vahedi V, Lieberman M A and Birdsall C K 1999 *Plasma Sources Sci. Technol.* **8** R45